\documentclass[12pt]{iopart}

\usepackage{iopams}
\usepackage{graphicx,bm}

\begin{document}

\title[]{Octahedral tilting, monoclinic phase and the phase diagram of PZT}

\author{F. Cordero$^{1}$, F. Trequattrini$^{2}$, F. Craciun$^{1}$ and C.
Galassi$^{3}$}
\address{$^1$ CNR-ISC, Istituto dei Sistemi Complessi,
Area della Ricerca di Roma -
Tor Vergata,\\
Via del Foss del Cavaliers 100, I-00133 Roma, Italy}
\address{$^{2}$ Dipartimento di Fisica, Universit\`{a} di Roma
\textquotedblleft La Sapienza\textquotedblright , P.le A. Moro 2,
I-00185 Roma, Italy}
\address{$^{3}$ CNR-ISTEC, Istituto di
Scienza e Tecnologia dei Materiali Ceramici, Via Granarolo 64,
I-48018 Faenza, Italy}

\begin{abstract}
Anelastic and dielectric spectroscopy measurements on PbZr$_{1-x}$Ti$_{x}$O$%
_{3}$ (PZT) close to the morphotropic (MPB) and antiferroelectric
boundaries provide new insight in some controversial aspects of its
phase diagram. No evidence is found of a border separating
monoclinic (M) from rhombohedral (R) phases, in agreement with
recent structural studies supporting a coexistence of the two phases
over a broad composition range $x<$ 0.5, with the fraction of M
increasing toward the MPB. It is also discussed why the observed
maximum of elastic compliance appears to be due to a rotational
instability of the polarisation and therefore cannot be explained by
extrinsic softening from finely twinned R phase alone, but indicates
the presence also of M phase, not necessarily homogeneous.

A new diffuse transition is found within the ferroelectric phase near $x
\sim $ 0.1, at a temperature $T_{\mathrm{IT}}$ higher than the well
established boundary $T_{\mathrm{T}}$ to the phase with tilted octahedra. It
is proposed that around $T_{\mathrm{IT}}$ the octahedra start rotating in a
disordered manner and finally become ordered below $T_{\mathrm{T}}$. In this
interpretation, the onset temperature for octahedral tilting monotonically
increases up to the antiferroelectric transition of PbZrO$_{3}$, and the
depression of $T_{\mathrm{T}}\left( x\right) $ below $x=0.18$ would be a
consequence of the partial relieve of the mismatch between the cation radii
with the initial stage of tilting below $T_{\mathrm{IT}}$.
\end{abstract}




\maketitle

\section{Introduction}

In spite of a prolonged and intensive research activity on the
ferroelectric perovskite PbZr$_{1-x}$Ti$_{x}$O$_{3}$ (also called
PZT $100\left( 1-x\right) /100x$) and related compounds, unsettled
issues remain on its phase diagram and on the microscopic mechanisms
that make it the most used piezoceramic material in many
applications. The highest electromechanical coupling is obtained at
compositions $x\lesssim 0.5$, near the morphotropic phase boundary
(MPB), which separates the rhombohedral (R) from the tetragonal (T)
region in the $x-T$ phase diagram \cite{JCJ71,Dam98}. For decades
such a high electromechanical coupling had been attributed to a
coexistence of R and T domains near the MPB, with consequent ease
for polarisation to change direction through domain switching or
domain wall motion. In 1999 Noheda \textit{et al.} \cite{NCS99}
found that below the MPB of PZT 52/48 the structure is monoclinic
(M), intermediate between R and T. This discovery stimulated
extensive investigations to better characterize such an intermediate
phase, where the direction of the polarisation can in principle
rotate continuously between the T $\left< 001 \right>$ and R $\left<
111 \right>$ directions; this provides an appealing explanation to
the enhanced ease for polarisation to change direction under an
external stress or electric field at the MPB. Yet, there is no
consensus on the nature and even existence of the M phase, with a
range of alternatives from its existence as uniform phase in a
narrow composition range, the coexistence with the R phase in a
broader range, or actually a fine mixture of R and T domains or
finely twinned R domains behaving on the average as monoclinic. The
debate has been recently reviewed \cite{Fra08b,PSB08,YZT09}, and
involves the mechanisms of rotation and switching of the
polarisation.

The situation is confused also on the Zr-rich end of the phase
diagram, where the ferroelectric R\ phase approaches the
antiferroelectric orthorhombic one. Here electron, but not neutron
or x-ray, diffraction experiments reveal superlattice peaks
incompatible with the R structure, variously attributed to rotations
of the O octahedra \cite{Vie95,VLD96} or to antiferroelectric-like
cation displacements away from the average $\left< 111 \right>$
direction \cite{RCW98,WKR05}.

Structural studies in these critical composition ranges face the
problem of analysing structures of domains with very short coherence
lengths and possibly of different coexisting phases, so that
information from other techniques, although not providing direct
information on the cell symmetry, may prove useful in clarifying
some issues. We present anelastic and dielectric spectroscopy
measurements, which provide new insight in these debates, and we
attempt to rationalize the phase diagram of PZT\ regarding the
tendency of the O octahedra to tilt.

\section{Experimental and Results}

The ceramic samples of PbZr$_{1-x}$Ti$_{x}$O$_{3}$, with Ti fractions $x=0.1$%
, 0.14, 0.17,\emph{\ }0.42, 0.45, 0.452,\ have been prepared similarly to a
previous study \cite{127} with higher values of $x$ (0.455, 0.465, 0.48 and
0.53), with the mixed-oxide method. The starting oxide powders were calcined
at 800~$^{\circ }$C for 4 hours (700~$^{\circ }$C for $x=0.1$), pressed into
bars and sintered at 1250~$^{\circ }$C for 2~h, packed with PbZrO$_{3}$\ +
5wt\% excess ZrO$_{2}$\ in order to maintain a constant PbO activity during
sintering. The powder X-ray diffractograms did not show any trace of
impurity phases and the densities were about 95\% of the theoretical ones.
The sintered blocks were cut into thin bars $4~$cm long and $0.6$~mm thick,
whose major surfaces were made conducting with Ag paint.

The dielectric susceptibility $\chi \left( \omega ,T\right) =\chi ^{\prime
}-i\chi ^{\prime \prime }$ was measured with a HP 4194 A impedance bridge
with a four wire probe and an excitation of 0.5 V/mm, between 0.2 and
500~kHz. The heating and cooling runs were made at $0.5-1.5$~K/min between
room temperature and 540~K in a Delta climatic chamber.

The mechanical analogue of the dielectric susceptibility is the elastic
compliance $s=s^{\prime }-is^{\prime \prime }$, which was obtained as the
reciprocal of the dynamic Young's modulus $E\left( \omega ,T\right)
=E^{\prime }+iE^{\prime \prime }=s^{-1}$. It was measured between 100 and
750~K by electrostatically exciting the flexural modes of the bars suspended
in vacuum on thin thermocouple wires \cite{135}. During a same run the first
three odd flexural vibrations could be tested, whose frequencies are in the
ratios $1:5.4:13.2$. The fundamental resonance angular frequency is \cite%
{NB72} $\omega \propto \sqrt{E^{\prime }}$, and the temperature variation of
the real part of the compliance is given by $s\left( T\right) /s_{0}\simeq $
$\omega _{0}^{2}/\omega ^{2}\left( T\right) $, where $\omega _{0}$ is chosen
so that $s_{0}$ represents the compliance in the paraelectric phase. The
imaginary parts of the susceptibilities contribute to the losses, which are
presented as $Q^{-1}=s^{\prime \prime }/s^{\prime }$ for the mechanical case
and $\tan \delta =\chi ^{\prime \prime }/\chi ^{\prime }$ for the dielectric
one.

\subsection{The new phase transformation at $T_{\mathrm{IT}}$ for $x\sim 0.1$%
}

\begin{figure}[tbp]
\includegraphics[width=8.5 cm]{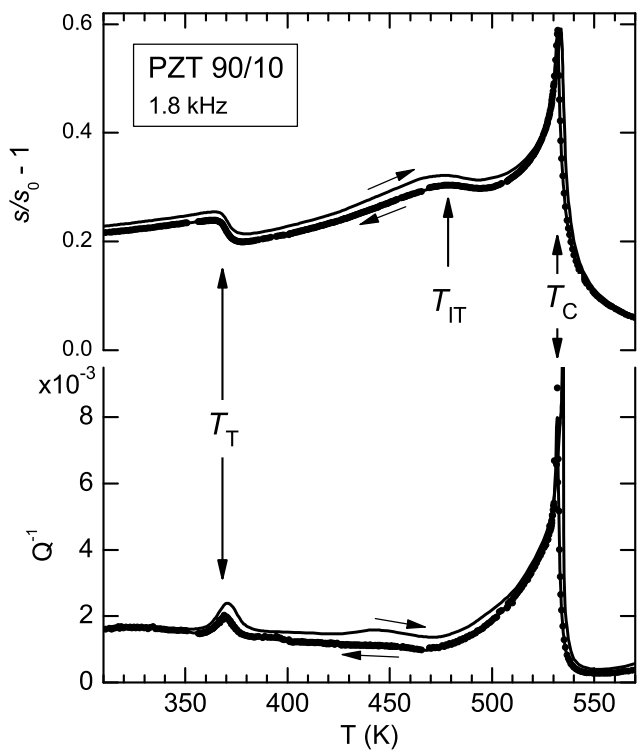} \vspace{-0 cm}
\caption{Anelastic spectrum of PZT 90/10 measured at 1.8~kHz during heating
and cooling.}
\label{fig an1hc}
\end{figure}

Figure \ref{fig an1hc} presents the anelastic spectrum of PZT 90/10. The
peak in both real and imaginary parts of the compliance at $T_{\mathrm{C}%
}=532.7$~K, with 1.6~K hysteresis between heating and cooling, signals the
transition between cubic paraelectric and ferroelectric phase, while the
anomaly at $T_{\mathrm{T}}=368$~K with an hysteresis of 1~K corresponds to
the transition from the rhombohedral $R3m$ (R) to the tilted $R3c$ (R$_{%
\mathrm{L}}$) phase, often labelled as R$_{\mathrm{H}}$ and
R$_{\mathrm{L}}$ respectively. These temperatures fall exactly on
the well known phase diagram of PZT, as shown in figure \ref{fig pd}
below. There is however an additional broad step in $s^{\prime }$,
without any counterpart in $Q^{-1}$, at the temperature
$T_{\mathrm{IT}}\simeq 477$~K, which we identify as the onset of an
intermediate tilt pattern of the octahedra, before the final tilt
pattern develops below $T_{\mathrm{T}}$.

\begin{figure}[tbp]
\includegraphics[width=8.5 cm]{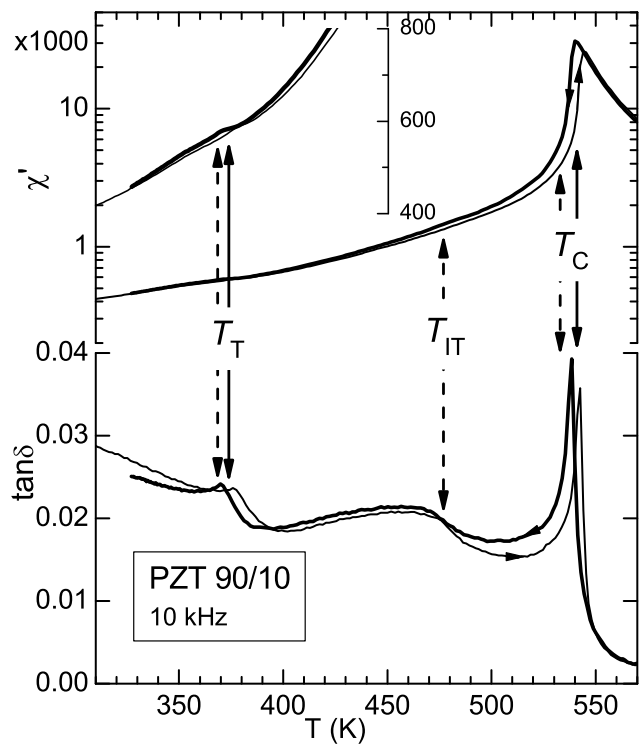} \vspace{-0 cm}
\caption{Dielectric spectrum of PZT 90/10 measured at 100~kHz during
heating and cooling. The dashed arrows are the transition
temperatures deduced from the anelastic spectra.} \label{fig diel}
\end{figure}

A signature of this new transition is present also in the dielectric
susceptibility as a broad step in $\tan \delta $ and a hardly
discernible hump in $\chi ^{\prime }$. This is shown in figure
\ref{fig diel}, together with the anomalies at $T_{\mathrm{C}}$ and
$T_{\mathrm{T}}$. The dashed arrows in figure \ref{fig diel} are the
transition temperatures deduced from the anelastic spectra in figure
\ref{fig an1hc}. The elastic compliance is peaked at a somewhat
lower temperature than the dielectric susceptibility at the
ferroelectric transition, as already observed \cite{FRM06}, and the
same is true for the tilt transition, but it should be noted that
since the two susceptibilities are differently coupled to the order
parameter, the shapes of their anomalies have to be different from
each other.

In order to check that there is indeed a structural transformation at $T_{%
\mathrm{IT}}$, we verified that the step in $s^{\prime }\left(
\omega ,T\right) $ is independent of the measuring frequency $\omega
/2\pi $, temperature rate and polarisation state. In fact, the
susceptibility curves may be affected by various processes,
especially in the presence of strong domain wall relaxations, as is
the case below the ferroelectric transition. When the temperature
rate is such that the domain configuration is kept far from
equilibrium, the susceptibility curve generally drops when the
temperature rate is decreased and partial aging proceeds. This
occurs also in certain temperature and composition ranges of PZT,
where marked irregularities may be induced in the $s\left( \omega
,T\right) $ curves by varying the temperature rate. We will not
discuss such phenomena, and only mention that they can be reduced by
keeping the temperature rate as constant and low as possible, and
their amplitude is larger at lower measuring frequency.

\begin{figure}[tbp]
\includegraphics[width=8.5 cm]{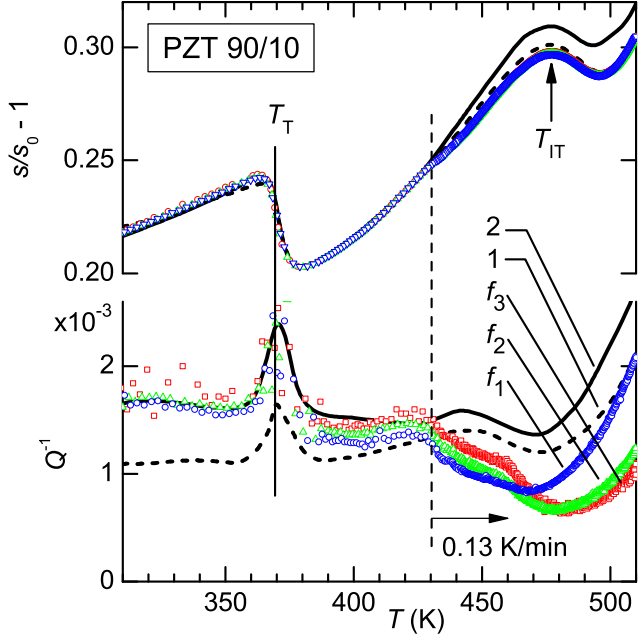} \vspace{-0. cm}
\caption{Anelastic spectra of PZT 90/10 measured during heating. 1 (dashed):
poled state, 1.8~kHz; 2 (solid): subsequent run (unpoled) at $3\pm 1$~K/min,
1.8 kHz; $f_{1}-$ $f_{3}$: subsequent run slowed down to 0.13~K/min above $%
\sim 430$~K (dashed vertical line) measured at $f=$ 1.8, 9.7 and
24~kHz.} \label{fig anRate}
\end{figure}

Figure \ref{fig anRate} shows that the anomaly at $T_{\mathrm{IT}}$
is instead perfectly reproducible and independent of all these
variables. Curve 1 (dashed) was measured in the initially polarised
state of the sample, obtained by application of 3~kV/mm~at
120~$^{\circ }$C for 40~min. The temperature rate was +1.3~K/min,
except near $T_{\mathrm{T}}$, where it was lowered to 0.8~K/min in
order to get enough accurate and closely spaced data points; the run
was extended up to 580~K, hence loosing the polarisation. Curve 2
(solid) is the subsequent heating run at $3\pm 1$~K/min. The
imaginary part is always above the previous run both because of the
higher density of relaxing domain walls in the unpoled state, and
because at the higher temperature rate such walls are more out of
equilibrium. Yet, the real parts are practically coincident until
$\sim 430$~K, where they start slowly departing from each other due
to the increasing influence of the ferroelectric domain wall
relaxation. The other three curves (empty circles) were measured
during the third heating run at $2-3$~K/min until $\sim 430$~K
and 0.13~K/min above that temperature; the frequencies are $f_{1}=1.8$~kHz, $%
f_{2}=9.7$~kHz and $f_{3}=24$~kHz. The important point is that both steps at
$T_{\mathrm{T}}$ and $T_{\mathrm{IT}}$ are practically the same in all the $%
s^{\prime }$ curves. This is true also for the peak in $Q^{-1}$ at $T_{%
\mathrm{T}}$, while the differences in the $Q^{-1}\left( \omega ,T\right) $
curves at higher temperature and particularly the drop when the rate is
lowered to 0.13~K/min have to be attributed to the tail of the domain wall
relaxations below $T_{\mathrm{C}}$. There is an additional peak or kink
around 445/460~K in the $Q^{-1}\left( \omega ,T\right) $ curves, which does
not appear to be connected with anomalies in the real part and will be
ignored. The lack of a peak in $Q^{-1}$ at $T_{\mathrm{IT}}$ and the
independence on frequency of the step in $s^{\prime }\ $also exclude that
the anomaly is caused by relaxation of any type of defects. Figure \ref{fig
anRate} therefore demonstrates that at the origin of the anomaly at $T_{%
\mathrm{IT}}$ there\ is a somewhat broadened but otherwise well behaved
phase transformation. We will consider $T_{\mathrm{IT}}$ as the temperature
of the upper edge of the rounded step in $s^{\prime }\left( T\right) $,
analogously to $T_{\mathrm{T}}$, where the assignment is corroborated by the
neat peak in $Q^{-1}\left( T\right) $.

\begin{figure}[tbp]
\includegraphics[width=8.5 cm]{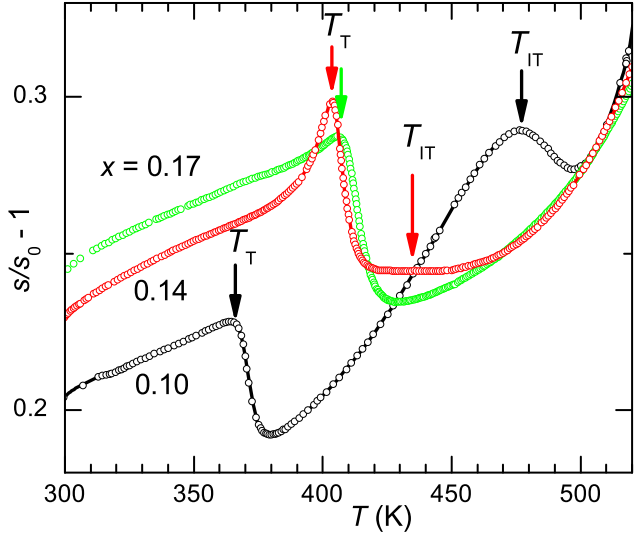} \vspace{-0 cm}
\caption{Anelastic spectra of PbZr$_{1-x}$Ti$_{x}$O$_{3}$ with $0.10
\leq x \leq 0.17$ measured at 1.8~kHz during heating.} \label{fig4}
\end{figure}

In order to draw a $T_{\mathrm{IT}}$\ line in the phase diagram, we also
measured the two concentrations $x=0.14$\ and 0.17, whose compliances are
plotted together with that of $x=0.10$\ in figure \ref{fig4}. The step at $T_{%
\mathrm{T}}$\ shifts to higher temperature according to the usual phase
diagram, though it acquires a peaked component, particularly evident at $%
x=0.14$, while the broader step at $T_{\mathrm{IT}}$\ shifts to lower
temperature merging with $T_{\mathrm{T}}$. That an anomaly exists for $%
x=0,14 $\ around 430~K is evident by comparing with the curve of $x=0.17$.
Both curves are practically coincident above 470~K and run parallel to each
other below the step at $T_{\mathrm{T}}$, but the $x=0.14$\ curve has a
clear bump indicated by the arrow, whose precise shape is however difficult
to evaluate. An even more attenuated anomaly might exist also for $x=0.17$,
even closer to $T_{\mathrm{T}}$, but it is not actually distinguishable.

\subsection{Compositions near the MPB}

\begin{figure}[tbp]
\includegraphics[width=8.5 cm]{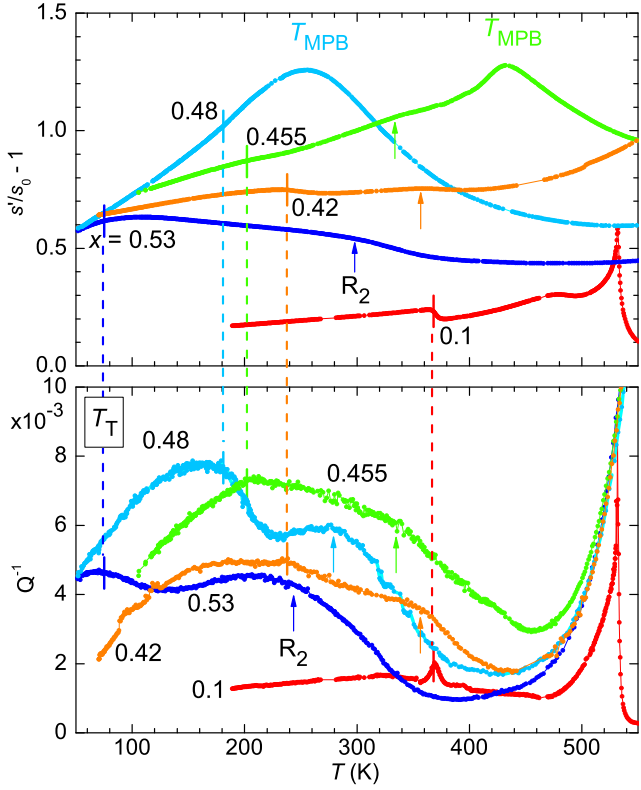} \vspace{0.2 cm}
\caption{Anelastic spectra of PbZr$_{1-x}$Ti$_{x}$O$_{3}$ with $x=0.1,$
0.42, 0.455, 0.48 and 0.53 measured at $1-1.8$~kHz. The vertical lines mark $%
T_{\mathrm{T}}$, while the arrows the relaxation R$_{2}$ (the three highest $%
x$ are from reference \cite{127}).} \label{fig TTvsx}
\end{figure}

The measurements on the samples at higher Ti content are similar to
those already published \cite{127} and a selection of them, including
some from reference \cite{127}, is presented in figure \ref{fig
TTvsx}. With increasing $x$, the anomaly at $T_{\mathrm{T}}$, marked
by vertical lines,\ becomes more diffuse and the spike in $Q^{-1}$
gradually transforms into a step of
increasing amplitude, at least until it occurs into the R/M phase, when $%
x<0.5$. This is consistent with the observation of a diffuse tilt transition
by neutron diffraction in PZT 60/40 \cite{ANC81,FHB08}. The step in $%
s^{\prime }$, instead, decreases its amplitude and becomes hardly
visible due to the peak at $T_{\mathrm{MPB}}$. The latter has been
attributed to the ability of the polarisation to continuously rotate
in the M phase between
the T and R directions \cite{127}. The analysis of the anomaly at $T_{\mathrm{%
T}}$ is also hindered by a broad peak in the losses around
$200-350$~K, labelled R$_{2}$ in reference \cite{127}, and whose
frequency dispersion indicates a relaxational origin rather than a
phase transformation. It appears also as a frequency-dependent hump
in $s^{\prime }$ and is indicated by arrows in figure \ref{fig
TTvsx} (the dependence on frequency is not shown). The curves with
$x=$ 0.45, 0.452, 0.455 and 0.465 are very similar to each other,
with $T_{\mathrm{T}}$ signaled by a weak step in $s^{\prime }\left(
T\right) $ and a small cusp in $Q^{-1}\left( T\right) $, which at
first appear as qualitatively different from the clear step in $Q^{-1}$ at $%
x=0.48$. The difference in the $Q^{-1}\left( T\right) $ curves, however, may
be less important than it appears, because one must take into account the
maximum R$_{2}$, which looses importance with respect to the spike/peak at $%
T_{\mathrm{T}}$ on approaching the MPB. It is therefore possible that, if
one were able to decompose the curves into the two contributions, R$_{2}$
and the anomaly at $T_{\mathrm{T}}$, the latter would result as a broadened
step already at $x\simeq 0.46$.

\section{Discussion}

\begin{figure}[tbp]
\includegraphics[width=8.5 cm]{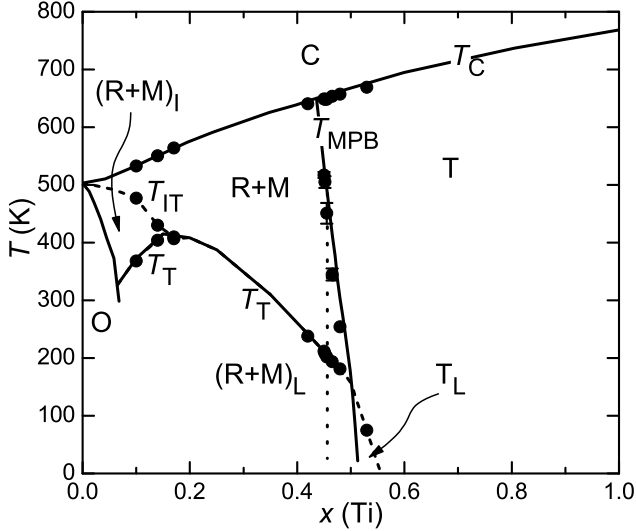} \vspace{-0. cm}
\caption{Phase diagram of PZT with the following phases: C paraelectric
cubic $Pm\bar{3}m$, O antiferroelectric orthorhombic $Pbam$; all the other
phases are ferroelectric: T tetragonal $P4mm$, T$_{\mathrm{L}}$ tilted
tetragonal $I4cm$, R+M mixed rhombohedral $R3m $ + monoclinic $Cm$, (R+M)$_{%
\mathrm{L}}$ mixed tilted rhombohedral $R3c$ + monoclinic $Cc$, (R+M)$_{%
\mathrm{I}}$ rhombohedral + monoclinic with intermediate tilt. The
filled circles are the transition temperatures deduced from our
measurements here and in reference \cite{127}. The dashed lines are
proposed here and the
dotted vertical line is the R/M boundary proposed by Noheda \protect\cite%
{NC06,NCS00}.} \label{fig pd}
\end{figure}

Figure \ref{fig pd}, presents the commonly accepted phase diagram of PZT \cite%
{JCJ71,NC06} (solid lines) together with the points deduced from our
anelastic spectra, including those of reference \cite{127}; the
dashed lines pass through our data and are drawn as explained in the
following paragraphs, but further measurements are necessary in
order to confirm their exact shape. The new anelastic and dielectric
experiments presented here contain essentially two results: the
existence of a novel phase transformation at $T_{\mathrm{IT}}$ below
$x\sim 0.17$ and the confirmation that there is a perfect continuity
of the $T_{\mathrm{T}}\left( x\right) $ line of the onset of
octahedral tilting up to the MPB, while $T_{\mathrm{T}}$ in the T
phase is lower than the extrapolation from the MPB.

The phases are the following: C is paraelectric cubic $Pm\bar{3}m$, O is
antiferroelectric (AFE) orthorhombic $Pbam$ with the octahedra rotated of
the same angle in anti-phase along the pseudocubic directions $\left[ 100%
\right] $ and $\left[ 010\right] $ ($a^{-}a^{-}c^{0}$ in Glazer's notation%
\cite{Gla72}) and antiferroelectric shifts of the cations along $\left[ 110%
\right] $ \cite{CGD97,WKR05}. All the other phases are ferroelectric
(FE), those just below the Curie temperature $T_{\mathrm{C}}$ having
unrotated octahedra: T is tetragonal $P4mm$ with polarisation along
$\left[ 001\right] $, R is rhombohedral $R3m$ with polarisation
along $\left[ 111\right] $, M is monoclinic $Cm$ with the
polarisation along a direction intermediate between T and R. Below
the $T_{\mathrm{T}}$ border the octahedra rotate
giving rise to R$_{\mathrm{L}}$ rhombohedral $R3c$ ($a^{-}a^{-}a^{-}$), T$_{%
\mathrm{L}}$ tetragonal $I4cm$ with tilt pattern $a^{0}a^{0}c^{-}$, as
predicted by first principle calculations \cite{KBJ06} and recently verified
by neutron diffraction at 5~K \cite{HSK10}, M$_{\mathrm{L}}$ monoclinic $Cc$
with tilt pattern $a^{-}a^{-}b^{-}$ intermediate between R$_{\mathrm{L}}$
and T$_{\mathrm{L}}$. The nature of the monoclinic phases, highly debated,
and of the new intermediate phase (R+M)$_{\mathrm{I}}$ are discussed next.

\subsection{MPB and search for the R/M border}

The phase diagram proposed by Noheda \textit{et al.}
\cite{NC06,NCS00}, and since then commonly adopted and reproduced
with first principle calculations \cite{KBJ06}, contains an almost
vertical border between the R and M phases
within the MPB range at $x_{\mathrm{R/M}}\simeq 0.455$ (dotted line in figure %
\ref{fig pd}), but the presence of this border is contradicted by recent
neutron diffraction experiments. The results of one of them \cite{FHB08} are
much better refined in terms of monoclinic $Cm$ rather than R structure at $%
x=0.4$, well within the supposed R region. In the other experiment \cite%
{YZT09} it is established that in ceramic samples the R and M phases
coexist over the whole composition range from the orthorhombic (O)
phase to the MPB, with the fraction of M phase increasing toward the
MPB. The actual relative fraction of the two phases, and some
details of the atomic displacements depend on the preparation
technique and a pure R phase can be obtained in single crystals,
only available at $x\leq 0.1$ \cite{YZT09,CG74,WCG78}. In addition,
there is considerable debate about the role of the short length
scale of the T, R and M domains near the MPB \cite{GTB04} and of the
high density of domain boundaries or microtwinning of the R domains,
that may constitute an adaptive phase with the average properties
attributed to the M phase
\cite{Vie00,JWK03,SSK07b,RW07,HSK10,Kha10}. First principle
calculations do not solve the issue, since in some studies a stable
monoclinic ground state is found \cite{ASC08} but in other cases is
excluded \cite{FFN07,FFZ09}.

In our data there is no sign of an R/M border, at least within the interval $%
0.42<x<0.48$. In fact, the $T_{\mathrm{T}}\left( x\right) $ line is
perfectly continuous within that interval, and the anelastic spectra
exhibit a smooth evolution from $x=0.42$ to $x=0.465$ and probably
also 0.48, as discussed in the explanation of figure \ref{fig
TTvsx}. Moreover, the anelastic spectra do not have any sign of an
additional M/R phase transition, expected if the M/R border existed
for $x>0.42$ and were not vertical, as in the theoretical phase
diagram obtained from first-principle calculations \cite{KBJ06}.

For all these reasons we labelled the region to the left of the MPB
as R+M, adopting the view of Yokota \textit{et al.} \cite{YZT09} of
a mixture of rhombohedral $R3m$ and monoclinic $Cm$ phases, both
with untilted octahedra. When discussing the $x=0.1$ composition,
however, we will for simplicity refer to the R phase only, which
certainly predominates over the M one. Of course, the absence of a
M/R boundary is natural in the hypothesis that the M phase is
actually the average of an adaptive microtwinned R phase.

\subsection{M versus R/T adaptive phase}

The different conclusions of the various analyses based on
thermodynamic approaches and first principle calculations probably
reflect the fact that the border between a true M phase and an
adaptive one is indeed vague. The MPB is the locus in the phase
diagram where the anisotropy of the free energy changes over from
stabilizing the R phase to the T phase. The main anisotropic term in
the Landau expansion in terms of powers of the polarisation
\textbf{P} is $\propto \left(
P_{1}^{4}+P_{2}^{4}+P_{3}^{4}\right) $ or its complementary $%
(P_{1}^{2}P_{2}^{2}+P_{2}^{2}P_{3}^{2}+P_{3}^{2}P_{1}^{2})$, with
extrema in the $\left\langle 100\right\rangle $ or $\left\langle
111\right\rangle $ directions, depending on the sign, which
stabilize the R or T phase. Including terms of the sixth order, the
anisotropy acquires extrema also along directions stabilizing the O
phase, while it is necessary to include eighth order terms in order
to stabilize the M phase \cite{VC01}. At the MPB the anisotropic
term of fourth order changes sign and therefore vanishes
\cite{FI97b,RKA08}: the material is isotropic with respect to
polarisation, at least up to the fourth order in \textbf{P,} and one
has a transversal or rotational instability of \textbf{P}
\cite{II98} accompanied by divergence of the transversal dielectric
susceptibility and some enhancement of the corresponding shear
compliance $s_{44}$ \cite{II99b}. A divergence of the compliance can
be obtained at the transition between T and O phase, including the
sixth order terms \cite{II99b}, but this is not the case of PZT,
which does not have a O phase at the MPB. The theoretical
expressions of the compliances from an expansion of the free energy
up to the eighth order are certainly cumbersome, but it is not
necessary to work them out in order to establish that a divergence
of $s_{44}$ is expected also at the T/M border. In fact, we argued
\cite{127} that in the case that a M phase exists in which
\textbf{P} can continuously tilt from the T to the R directions with
little change of the magnitude, the rotation angle from the original
T direction acts as order parameter \cite{Hud08} and is almost
linearly coupled to the strains of symmetry $\Gamma _{5-}$
($\varepsilon _{4}$, $\varepsilon _{5}$ and $\varepsilon _{6}$ in
Voigt notation), thereby causing a divergence in the respective
compliance $s_{44}$ \cite{Reh73,Sal90}. Accordingly, we interpreted
the maximum in the compliance at the MPB, particularly pronounced at
$x=0.465$, as due to this rotational instability in the M phase
\cite{127}.

The main arguments against a true M phase are that it is very
unlikely that PZT is so anharmonic to require terms up to the eight
power of \textbf{P} in the expansion of the free energy
\cite{Kha10,RKA08}. As a consequence, the real structure should be
R, but with fine twinning thanks to the near isotropy at the MPB,
which lowers the energy of the domain walls. Such an adaptive phase,
analogous to the heavily twinned phases in martensites, would
produce diffraction patterns easily confused with a homogeneous M
phase and would also cause extrinsic softening \cite{Kha10,RKA08}.
The first part of the argument is certainly convincing for an ideal
homogeneous crystal, but PZT has both quenched strain fields from
the cation disorder \cite{VC01} and also internal fields from the
domain configuration, which presents particularly wavy and strained
walls in the R phase \cite{SSK07b}. While these microstructure and
internal strains may let the diffraction patterns of the R phase
appear as an homogeneous M phase, they also modify the local
anisotropy, acting as higher order terms in the free energy
expansion and stabilizing an M phase \cite{VC01}. We would therefore
expect that there are regions where a real M phase is formed,
although with cell parameters and direction of the polarisation
dependent on the local field. A border between a mixed M+R
\cite{YZT09} and a polar glass state \cite{Kha10} is probably
impossible to establish, and the balance between the two
descriptions may also depend on the sample preparation and
microstructure, not to mention whether the sample is bulk ceramic or
thinned for electron microscopy.

Here we would like to point out an aspect that distinguishes between
M, not necessarily homogeneous, and microtwinned adaptive R phase
and is overlooked in the debate: the dynamics. In fact, an adaptive
phase with high density of domain walls having vanishing energies
causes extrinsic softening \cite{Kha10,RKA08}, but we are not aware
of a detailed theory of its frequency and temperature dependence. We
would expect an important contribution of relaxational nature, with
a consequent dependence on frequency, and possibly also nonlinear
response, none of which we observe \cite{127}. Certainly a
complicated and fine domain structure exists, and it may well be at
the origin of the nonlinear stress-strain response of PZT
\cite{Kha10,RPZ06}, but the elastic response to low amplitude
excitation that we measure is strictly linear and almost frequency
independent. Again, the border between extrinsic softening from fine
twinning and intrinsic linear softening due to a rotational
instability may be blurry, since the vanishing of the orientational
free energy barrier brings about a crossover from thermally
activated, characteristic of domain walls, to almost athermal
dynamics, but the peak in $s^{\prime }$ at $T_{\mathrm{MPB}}$ is
more characteristic of the latter.

\subsection{The octahedral tilt transition}

From the present experiments we are not able to establish whether
the tilting transition affects both M and R\ domains in the same
manner or not, because we are not able yet to interpret the
evolution of the shape and amplitude of the anomaly at
$T_{\mathrm{T}}$ with varying $x$ (figure \ref{fig TTvsx}), whose
analysis is also made difficult by the relaxation R$_{2}$ (figure
\ref{fig TTvsx}).

We would like to stress the different origin of the hump corresponding to R$%
_{2}$ with respect to the anomalies at $T_{\mathrm{T}}$ and $T_{\mathrm{IT}}$%
, since they all appear similar in the $s^{\prime }\left( T\right) $ curves,
and it may be tempting to identify R$_{2}$ as another structural transition,
for example tilts in the M phase instead of the R phase. A more careful
examination, however, indicates that this is not the case. The anomalies at $%
T_{\mathrm{T}}$ and $T_{\mathrm{IT}}$ have all the characteristics
of well behaved, though diffuse, structural transformations, as
described above and in reference \cite{127} for $T_{\mathrm{T}}$:
they are perfectly reproducible during heating and cooling, and are
independent of frequency and temperature rate. Instead, R$_{2}$ has
none of these characteristics: its amplitude is larger at lower
frequency and strongly depends on the temperature rate (data not
shown here), so that we attribute it to kinetic effects of the
rearranging ferroelectric domain walls or adaptive microtwins,
rather than to a phase transformation. A\ final argument against the
involvement of octahedral tilts in the relaxation R$_{2}$ is that it
is
observed also in the T phase just below $\sim 300$~K (curve $x=0.53$ in figure %
\ref{fig TTvsx}), where tilting has never been observed.

The tilted tetragonal phase is predicted to have $I4cm$\ space
group, based on first-principle calculations \cite{KBJ06} and recent
neutron diffraction observations \cite{HSK10}. A cell doubling
transition in the T\ phase, possibly due to an octahedral anti-phase
rotation, was proposed by Ragini \textit{et al.} \cite{RMP01}, after
the observation in PZT 48/52 of superlattice peaks below 189~K and
anomalies in the dielectric $\chi ^{\prime }$\ and resonance
frequency, but the temperature of such anomalies is definitely
higher than the $T_{\mathrm{T}}$\ we found at the same composition
\cite{127}. We cannot exclude that the technique of sample
preparation has an influence on the position of the
$T_{\mathrm{T}}\left( x\right) $\ line in the high $x$\ region,
considering that it may affect the fraction of R and M phases
\cite{YZT09}.

In figure \ref{fig pd} we drew the $T_{\mathrm{T}}\left( x\right) $\
line straight until the MPB and then again straight with a higher
slope in order to interpolate the only available point in the T
phase (dashed line). Such a kink at the MPB is not actually
observed; it might be either smoother or a more discontinuous step
and future work is necessary in order to clarify this point.

\subsection{$T_{\mathrm{T}}\left( x\right) $ border and tolerance factor}

The occurrence of tilting transitions of the BO$_{6}$ octahedra is an
extremely common phenomenon in ABO$_{3}$ perovskites, and it has been
rationalized in terms of the tolerance factor \cite{Att01,Goo04}%
\begin{equation}
t=\frac{r_{\mathrm{A}}+r_{\mathrm{O}}}{\sqrt{2}\left( r_{\mathrm{B}}+r_{%
\mathrm{O}}\right) }~,
\end{equation}%
which is 1 if the mean ionic radii, usually taken from Shannon's tables \cite{SP69},
exactly match the A-O and B-O lengths in a cubic cell. When $t<1$,
the B-O bond is too long with respect to the A-O one; hence the A-O-A
network exerts a compression over the network of octahedra, which rotate in
order to accommodate the mismatch while keeping the B-O bonds long. The A-O
bonds are $\sqrt{2}$ longer and therefore are weaker and have larger thermal
expansion than the B-O bonds; therefore $t$ always decreases on cooling, and
when it drops below a critical value, within the interval $0.97<t<1$ for
most perovskites \cite{LBW06}, a tilting transition occurs. The tolerance
factor can also be decreased in a solid solution by increasing the mean
ionic radius in the B\ sublattice or decreasing that in the A sublattice.
Moreover, a sequence of tilting transitions may occur with proceeding
cooling and/or changing of the ionic radii, usually through more symmetrical
tilt patterns first, like rhombohedral $a^{-}a^{-}a^{-}$, and then to the
more distorted orthorhombic patterns \cite{Att01,Goo04}. This framework is
widely adopted to explain the phase diagrams of all types of perovskites
with chemical substitutions, including the ferroelectric ones \cite%
{WKR05,RCS94} and even in the presence of O vacancies \cite{140},
and PZT is no exception. The Shannon radii of Ti and Zr in
octahedral coordination are 0.605 and 0.72~\AA , respectively,
yielding $t=1.027$ for PbTiO$_{3}$, which indeed undergoes below
$T_{\mathrm{C}}$ off-centering of Ti in too large and
untilted octahedra, to $t=0.97$ of PbZrO$_{3}$, which undergoes a $%
a^{0}b^{-}b^{-}$ tilt transition with AFE shifts of the cations \cite{KWR03},
having passed through the more symmetrical rhombohedral $a^{-}a^{-}a^{-}$
structure at intermediate compositions. Conforming to the above discussion,
the $T_{\mathrm{T}}\left( x\right) $ line encloses the region of low
tolerance factor, namely low $T$ and low $x$, possibly with some
discontinuity when passing from the T to the R+M region, from $x>0.5$ down
to $x=0.18$. At this point it does not prosecute to the maximum value at $%
x=0 $, but decreases sharply. The departure from the expected monotonic rise
has been already noted and explained in terms of competition between the O
AFE phase with $a^{-}a^{-}c^{0}$ tilt and the R FE phase; this would cause a
frustration of the Pb displacements and consequently inhibit octahedral
tilting, which is coupled with such displacements \cite{WKR05}.

The presence of a phase transition at a $T_{\mathrm{IT}}\left( x\right) $\
merging with $T_{\mathrm{T}}\left( x\right) $\ around $x\sim 0.17$\
alternatively suggests that the phase (R+M)$_{\mathrm{I}}$, in addition to
the cation shifts away from $\left[ 111\right] $, has an initial stage of
octahedral tilting. In other words, if a structural transition occurs in
that temperature range, it seems more likely that its driving force acts on
the octahedral tilting rather than on antiferroelectric cation shifts. In
this manner the depression of the boundary of the $R3c$ phase is a
consequence of the fact that the mismatch between (Ti/Zr)-O bonds and Pb-O\
bonds has been relaxed by the first tilting transition at $T_{\mathrm{IT}}$,
and further cooling below $T_{\mathrm{T}}$ is necessary in order to trigger
the final $a^{-}a^{-}a^{-}$ pattern. In this view, the frustration between
the FE and AFE phases would mainly broaden and split in two stages the
tilting transition, and the depression of the combined $T_{\mathrm{IT}}-T_{%
\mathrm{T}}$\ line near $x=0.17$\ is much less important than that
of the $T_{\mathrm{T}}$\ curve alone. The existence of coupling
between tilts and polar modes is recognizable by the effect of the
transition at $T_{\mathrm{T}}$ on the polarisation
\cite{GMC78,CNI97}, and dielectric susceptibility \cite{VLD96} (see
also the present data), by the reduction of the extrinsic
contribution to piezoelectric effect in the tilted $R3c$ phase
\cite{ER07} and is also indicated by first-principle calculations
\cite{LCW02}.

\subsection{Nature of the intermediate phase below $T_{\mathrm{IT}}$}

We can only make conjectures on the nature of the transition at $T_{\mathrm{%
IT}}$ based on our susceptibility experiments. If it is indeed an
initial stage of the rotations of the octahedra, it may be a tilt
pattern intermediate between those of the O and R$_{\mathrm{L}}$
phases, like $a^{-}a^{-}b^{-}$ \cite{WKR05}, or it may involve the
$M_{3}$ in-phase rotation modes \cite{VLD96,BBV10}. Indeed, although
no experimental phonon-dispersion data are available for PbZrO$_{3}$
\cite{TTO09}, first-principles calculations indicate that the whole
$R_{25}-M_{3}$ branch is unstable, and not only the $R_{25}$ mode of
anti-phase rotations \cite{GCW99}. It is therefore possible that a
combination of the two types of instabilities produces a relatively
disordered tilt pattern, which becomes the ordered $a^{-}a^{-}a^{-}$
($R3c$) or $a^{-}a^{-}b^{-}$ ($Pc$) structure below
$T_{\mathrm{T}}$. The concept of disordered tilts has been proposed
for explaining why the techniques providing snapshots of the local
structure, the X-ray absorption spectroscopies EXAFS and XANES and
the pair distribution function from neutron diffraction, do not see
any change at the tilt transitions of perovskites like NaTaO$_{3}$
\cite{RYS94} and the same PZT \cite{TEV96}. The situation is similar
for most of the apparently displacive transformations in perovskites
involving off-centering of cations, like the FE transitions in
BaTiO$_{3}$ and PbTiO$_{3}$. The issue has been solved by assuming
that those transitions have an important order-disorder component,
so that the Ti atoms are off-center also in the cubic phase, but
without correlation between different cells \cite{SRY94,FFP99}. The
application of the same concept to the transitions involving
concomitant rotations of the octahedra about more than one direction
is less obvious, since completely disordered tilts would require
excessive distortions of the octahedra. Yet, the idea that the phase
$R3m$ of PZT is untilted on the average but locally tilted has been
adopted by other authors \cite{Vie95,VLD96}, also to justify the
fact that first principle calculations indicate the tilted $R3c$
structure and not the $R3m$ one as the ground state of PZT
\cite{LCW02}. In our case, the hypothesis that the initially
disordered tilting below $T_{\mathrm{IT}}$ becomes long range
ordered below $T_{\mathrm{T}}$ provides a rationale for the
different aspects of the two anomalies:\ the one at
$T_{\mathrm{IT}}$ is diffuse due to the dynamical frustration of the
tilts, while that at $T_{\mathrm{T}}$ is sharp.

Although the transition at $T_{\mathrm{IT}}$ had never been reported
before, the existence of an additional phase of still controversial
nature in the PZT phase diagram near the boundary to the O phase is
indicated by electron diffraction experiments. In such experiments,
$\frac{1}{2}\left\{ kl0\right\} $ reflections have been observed
near room temperature, which are incompatible with both the O and R
phases, and have first been attributed to in-phase tilting of the
octahedra \cite{Vie95}. Later, it has been argued that rotations or
distortions of the O octahedra would not be sufficient to produce
spots with the observed intensities \cite{RCW98}; moreover, in-phase
rotations would not produce reflections with $h=k$, which are
instead observed, and are generally unlikely to occur in perovskites
with $t$ only slightly smaller than 1 \cite{WKR05}. The superlattice
reflections have therefore been attributed to antiferroelectric
shifts of Pb away from $\left< 111\right> $ in the $\left<
110\right> $ directions, as in the neighboring AFE O phase. A
difficulty in characterizing the new structure is that the
superlattice reflections are observed only in electron diffraction,
but not in X-ray and neutron diffraction \cite{RCW98}; possible
reasons are small scattering intensities due to the different cross
sections in the different techniques, and the occurrence of these
atomic displacements as surface effects in the very thin samples
used for TEM \cite{RCW98}. Nonetheless, a narrow region of
intermediate phase between the AFE O and the FE R phases has been
proposed \cite{WKR05}, untilted $Pm$ at higher temperature and
tilted $Pc$ below the usual $T_{\mathrm{T}}$ border. According to
this phase diagram, $T_{\mathrm{IT}}$ would correspond to the
transition from $R3m$ to $Pm$.

While the present measurements on ceramic samples prove that the presence of
an intermediate phase below $T_{\mathrm{IT}}$ is a bulk phenomenon, we feel
that the question whether the transition involves only cation shifts or also
octahedral rotations is yet open, in view of the expected small intensity of
reflections connected with O shifts \cite{RCW98}. The issue should be
clarified by careful neutron diffraction experiments, which are more
sensitive to O displacements and do not involve the uncertainties connected
with very thin samples where surface effects are important and the actual
temperature under the electron beam may be considerably higher than expected.

\subsection{Signatures of the transition at $T_{\mathrm{IT}}$ in the
previous literature}

There is abundant literature on PZT near the $x=0.1$ composition, so that it
seems appropriate to examine possible signatures of the transition at $T_{%
\mathrm{IT}}$ in previous studies, which had been overlooked because smaller
and less definite than the anomalies at $T_{\mathrm{T}}$.

A first indication of a transition at $T_{\mathrm{IT}}$ comes from a
minor step in the steeply falling $P\left( T\right) $\ curve of PZT
90/10 in figure 6 of reference \cite{WCG78}, although it had not
been recognized as such, being less clear than the transition at
$T_{\mathrm{T}}$. Also the temperature dependence of the
rhombohedral angle $\alpha \left( T\right) $\ in PZT 90/10 may be
re-examined. Such an angle is related to the octahedral tilt angle
in the R structure \cite{MMG70}, and the lack of clear change of
slope of $\alpha \left( T\right) $\ at $T_{\mathrm{T}}$\ was
explained \cite{CG74} as due to dynamic fluctuations of the tilt
angle above $T_{\mathrm{T}}$, an explanation very similar to the
assumption of a $R_{\mathrm{I}}$
phase of disordered tilts. The difference is that the tilt disorder in the $%
R_{\mathrm{I}}$ phase would not be dynamic but almost static with onset
around $T_{\mathrm{IT}}$.

The fact that the local symmetry of PZT 90/10 is lower than
rhombohedral is confirmed by recent infrared and Raman spectroscopy
experiments \cite{BBV10}, where several additional modes are
observed, besides those of a uniform R phase. Such modes are
compatible with the additional disordered Pb shifts in the
orthorhombic directions \cite{RCW98,WKR05} but also with the $M_{3}$
in-phase tilt mode of the O octahedra \cite{BBV10}, which, together
with the anti-phase $R_{25}$ mode, would produce a disordered tilt
pattern. In the same study, dielectric susceptibility curves of PZT
90/10 are shown, which are very similar to those of figure \ref{fig
diel}, except for the lack of a clear anomaly at $T_{\mathrm{IT}}$;
the hump in the imaginary part, however, appears with 2\% La
doping \cite{BBV10}.

\section{Conclusions}

New anelastic and dielectric spectroscopy experiments are presented on PbZr$%
_{1-x}$Ti$_{x}$O$_{3}$ with $0.1\leq x\leq 0.17$\ and $0.42\leq $ $x\leq
0.452$, providing new information on the phase diagram of PZT at the
critical compositions near the antiferroelectric and morphotropic borders.
At $x=0.1$, a new diffuse phase transformation is found at $T_{\mathrm{IT}%
}\sim 480$~K, which is interpreted as the onset of disordered tilting of the
O octahedra, before the long-range ordered tilt pattern develops at the well
established sharp transition at $T_{\mathrm{T}}=368$~K. Indications of such
a phase transformation from earlier literature are also reported. On
increasing Ti content, $T_{\mathrm{IT}}$ decreases and merges with $T_{%
\mathrm{T}}$ around $x\sim 0.17$. In this interpretation, the onset
temperature for octahedral tilting monotonically increases from $T_{\mathrm{T%
}}=0$ within the Ti-rich tetragonal phase up to the antiferroelectric
transition of PbZrO$_{3}$. The well known depression of $T_{\mathrm{T}%
}\left( x\right) $ below $x=0.18$ would be due to the fact that the mismatch
between the cation radii is partially relieved at the initial transition at $%
T_{\mathrm{IT}}$. The combined $T_{\mathrm{IT}}-T_{\mathrm{T}}$\
border presents a much shallower depression near $x=0.17,$ but
further anelastic and dielectric experiments at additional
compositions with $x<0.25$ are necessary to clarify the exact shape
of the borders of the phase diagram, for example if the
$T_{\mathrm{IT}}$ line merges with $T_{\mathrm{T}}$, as depicted in
figure \ref{fig pd}, or crosses it and prosecutes, as proposed by
Woodward \textit{et al.} \cite{WKR05} Yet, the exact nature of the
intermediate phase or phases close to the border with the AFE O
phase should be clarified by additional diffraction experiments,
analysed with the consciousness of the existence of a transition at
$T_{\mathrm{IT}}$.

It is discussed why the frequency independent maximum of the linear
and elastic compliance at the MPB can hardly be explained by finely
twinned R phase alone, and rather indicates the presence of a
monoclinic phase, possibly stabilized by the internal random strains
and coexisting with the rhombohedral one. The $T_{\mathrm{T}}\left(
x\right) $ line is also shown to be perfectly continuous up to the
MPB, which, together with a smooth evolution of the anelastic
spectra with varying composition, provides evidence against a clear
border between rhombohedral and monoclinic phases, and rather
confirms the recent structural studies where the two structures are
found to coexist over a broad composition range.

\ack

The authors thank Mr. C. Capiani (ISTEC) for the skillful preparation of the
samples, Mr. F. Corvasce (ISC), P.M. Latino (ISC) and A. Morbidini (INAF)
for their technical assistance in the anelastic and dielectric experiments.

\end{document}